\documentclass[journal=nalefd,layout=twocolumn,manuscript=letter]{achemso}
\usepackage[T1]{fontenc}
\usepackage{amsmath}
\usepackage{amssymb}
\usepackage[flushleft]{threeparttable}
\usepackage{graphicx}
\usepackage{bm}
\usepackage{comment}
\usepackage[hidelinks]{hyperref}
\usepackage{braket}
\usepackage{color}
\usepackage{array}
\usepackage{siunitx}
\usepackage{todonotes}

\author{Barbara Ursula Lehner}%
\affiliation{Institute of Semiconductor and Solid State Physics, Johannes Kepler University, Linz, 4040, Austria}
\alsoaffiliation{Secure and Correct Systems Lab, Linz Institute of Technology, 4040 Linz, Austria}
\email{barbara.lehner@jku.at}
 
\author{Tim Seidelmann}%
\affiliation{Theoretische Physik III, Universität Bayreuth, 95440 Bayreuth, Germany}

\author{Gabriel Undeutsch}%
\affiliation{Institute of Semiconductor and Solid State Physics, Johannes Kepler University, Linz, 4040, Austria}

\author{Christian Schimpf}%
\affiliation{Institute of Semiconductor and Solid State Physics, Johannes Kepler University, Linz, 4040, Austria}

\author{Santanu Manna}%
\affiliation{Institute of Semiconductor and Solid State Physics, Johannes Kepler University, Linz, 4040, Austria}

\author{Michał Gawełczyk}%
\affiliation{Department of Theoretical Physics, Faculty of Fundamental Problems of Technology, Wroc\l{}aw University of Science and Technology, 50-370 Wroc\l{}aw, Poland}

\author{Saimon Filipe Covre da Silva}%
\affiliation{Institute of Semiconductor and Solid State Physics, Johannes Kepler University, Linz, 4040, Austria}

\author{Xueyong Yuan}%
\affiliation{School of Physics, Southeast University, Nanjing 211189, China}

\author{Sandra Stroj}
\affiliation{Forschungszentrum Mikrotechnik, FH Vorarlberg, 6850 Dornbirn, Austria}

\author{Doris E. Reiter}%
\affiliation{Condensed Matter Theory, TU Dortmund, 44221 Dortmund, Germany}

\author{Vollrath Martin Axt}%
\affiliation{Theoretische Physik III, Universität Bayreuth, 95440 Bayreuth, Germany}

\author{Armando Rastelli}%
\email{armando.rastelli@jku.at}
\affiliation{Institute of Semiconductor and Solid State Physics, Johannes Kepler University, Linz, 4040, Austria}

\title{{Beyond the four-level model: Dark and hot states in quantum dots degrade photonic entanglement}}

\abbreviations{QD, LDE, SPDC, FSS, DBR, TPE, PL, X, XX}
\keywords{Quantum Optics, Quantum Dots, Temperature Dependency, Excited States, Hot States, Entanglement}

\begin{document}
\begin{abstract}
Entangled photon pairs are essential for a multitude of photonic quantum applications. To date, the best performing solid-state quantum emitters of entangled photons are semiconductor quantum dots operated around liquid-helium temperatures. To favor the widespread deployment of these sources, it is important to explore and understand their behavior at temperatures accessible with compact Stirling coolers. Here we study the polarization entanglement among photon pairs from the biexciton-exciton cascade in GaAs quantum dots at temperatures up to $\sim\SI{65}{K}$. We observe entanglement degradation accompanied by changes in decay dynamics, which we ascribe to thermal population and depopulation of hot and dark states in addition to the four levels relevant for photon pair generation. Detailed calculations considering the presence and characteristics of the additional states and phonon-assisted transitions support the interpretation. We expect these results to guide the optimization of quantum dots as sources of highly entangled photons at elevated temperatures. 
\end{abstract}

\section{Introduction}
\begin{figure*}[htb]
	\includegraphics[width=1\textwidth]{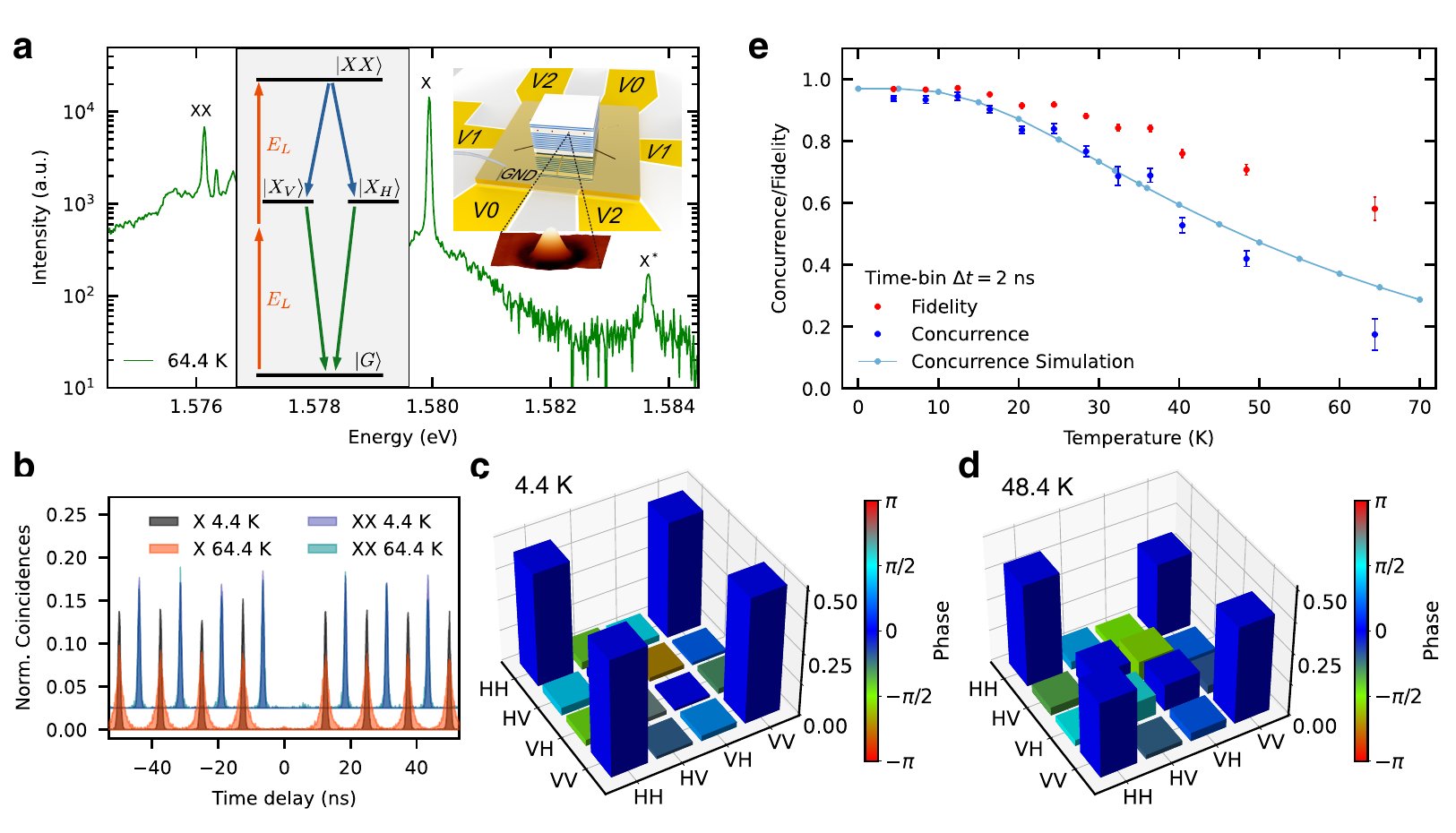}
	\caption{Measurements on a GaAs quantum dot at temperatures $T$ up to about 65~K. (a) Photoluminescence spectrum under resonant two-photon excitation at nominal $\SI{64.4}{K}$. Besides the exciton (X) and biexciton (XX), radiative transitions of higher-energy excitonic states X$^*$ become visible. Laser stray light is overlaid by the excitation scheme showing the laser energy $E_L$ and the basic 4-level model including biexciton $\ket{XX}$, bright excitons $\ket{X_{H/V}}$, and the ground state $\ket{G}$. Top right: Sketch of the strain-tuning device with the sample containing the QDs and $150\times150\times\SI{9}{nm^3}$ AFM image of a GaAs QD. (b) $g^{(2)}$ auto-correlation measurements of X and XX at $\SI{4.4}{K}$ and $\SI{64.4}{K}$. (c) and (d) Two-photon density matrices reconstructed from polarization-resolved cross-correlation measurements at $\SI{4.4}{K}$ and $\SI{48.4}{K}$. (e) Measured concurrence (blue dots) and fidelity (red dots) by using a $\SI{2}{ns}$ time bin. The theoretical simulation of the concurrence is shown in light blue.}
	\label{fig:fig1}
\end{figure*}

Entangled photon pairs have been used to explore the validity of quantum mechanics and some of its least intuitive predictions~\cite{Bouwmeester1997}. Besides its appeal for fundamental research, entanglement is a key resource to establish correlations among remote locations, to achieve resolution beyond classical capabilities and for quantum information processing\cite{Knill2001,OBrien2009,Pan2012}. In the last decades different approaches have been developed to generate entangled photon pairs~\cite{Orieux2017}.
The most established methods are based on spontaneous parametric down-conversion (SPDC)~\cite{Kwiat1995SPDC,Kwiat1999}, which has led to sources that can be operated in a wide temperature range~\cite{Pan2021hightempSPDC} and also in satellites~\cite{Yin2020}. However, the stochastic generation of photons leads to an increase of the multi-pair generation probability and thus to a degradation of entanglement~\cite{Joens2017,Schneeloch_2019} when the brightness is increased. In contrast, semiconductor quantum dots (QDs)\cite{Senellart2017} are quantum emitters exhibiting sub-Poissonian emission characteristics and ultra-low multi-photon pair emission probability, independent of the excitation rate. In ideal QDs, a polarization-entangled photon pair can be obtained by initializing the system in a biexciton ($\ket{XX}$) state~\cite{Benson2000,Huber2018a}, which decays back to the crystal ground state $\ket{G}$ via two bright and energy-degenerate excitonic ($\ket{X_{H/V}}$) states following two possible decay paths, as sketched in the left inset of~\autoref{fig:fig1}(a). In particular, GaAs QDs obtained via the local droplet etching (LDE) method in an AlGaAs matrix \cite{Gurioli2019,Saimon2021,Wang2007,Heyn2009,chen2018} have demonstrated excellent performance as sources of single entangled photon pairs with a fidelity to one of the maximally entangled Bell states as high as 0.98 \cite{Huber2018,Schweickert2018}. Thus far the best results have been obtained at cryogenic temperatures, reachable with liquid-helium-based (wet) cryostats or bulky and energy-intensive closed-cycle (dry) cryostats.
To achieve further advances for QD light sources, possibly enabling their deployment in space applications with compact and energy-efficient cryo-coolers~\cite{Schlehahn2015}, a comprehensive study of entanglement at different temperatures is needed. By using strain-tunable GaAs QDs capable of generating nearly perfectly entangled photons at low temperatures~\cite{Huber2018}, we investigate the effects produced by increasing operation temperature $T$ on the entanglement and on the exciton decay dynamics following coherent $\ket{XX}$ excitation. While the multi-pair emission probability remains low for all investigated temperatures, we find that entanglement degrades for $T\gtrsim\SI{15}{K}$, which is accompanied by a slowed excitonic decay and weak light-emission from hot excitonic states. To gain insight into this rich evolution we (i) expand the four-level model for the biexciton-exciton cascade by including higher-energy (``hot'') and dark excitonic states, (ii) address the properties of such additional levels as well as of the corresponding radiative and non-radiative transitions by experiments and 8-band $k\cdot p$ and configuration-interaction (CI) calculations, (iii) model the population dynamics with the Liouville-von-Neumann equation with Lindblad terms and rate-equations, and (iv) evaluate the two-time correlation functions and degree of entanglement based on density matrix methods. Our calculations reproduce very well the experimental results and provide evidence that both the degradation of entanglement and the changes in decay dynamics for the $\ket{XX}\rightarrow\ket{X_{H/V}}$ (shortly XX) and $\ket{X_{H/V}}\rightarrow\ket{G}$ (X) transitions can be traced back to the thermal population and relaxation of excited and dark exciton states and to spin mixing.
\begin{figure*}[htbp]
	\includegraphics[width=1\textwidth]{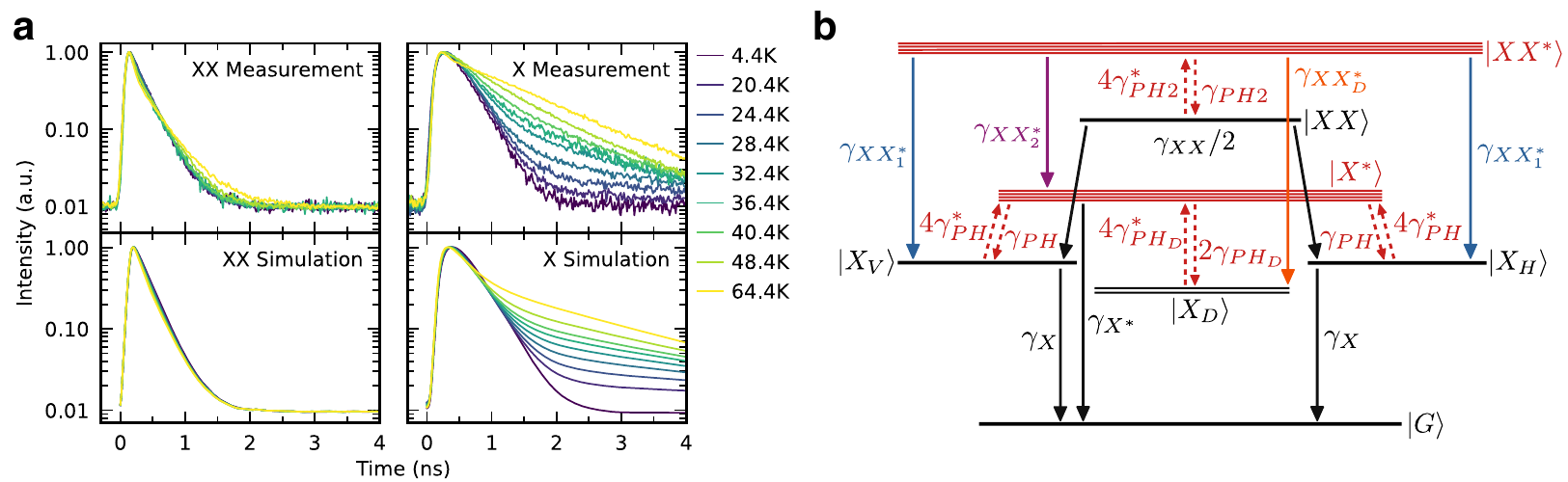}
	\caption{(a) Measurement and simulation of biexciton (XX) and exciton (X) decay dynamics at temperatures from $\SI{4.4}{K}$ to $\SI{64.4}{K}$. (b) Sketch of the energy levels used to model the decay dynamics. Excited biexciton and exciton states ($\ket{XX^*}$ and $\ket{X^*}$) are shown in red, while the four levels relevant for the generation of entangled photon pairs as well as the dark excitonic states are shown in black. $\gamma_X$ $\left(\gamma_{XX}\right)$ is the rate associated with the radiative decay of the $\ket{X}$ $\left(\ket{XX}\right)$ states. Temperature-dependent transitions due to phonon emission and absorption processes are shown as dashed red arrows, whereby the prefactors correspond to the multiplicity of the $\ket{XX^*}$, $\ket{X^*}$ and the dark exciton $\ket{X_D}$ states, assumed energetically degenerate. The rates of the shown transitions can be found in \autoref{tab:tab1}.}
	\label{fig:fig2}
\end{figure*}
\section{Results}

GaAs QDs grown via the LDE method are nearly strain-free and their properties can be optimized so that the bright excitonic states $\ket{X_{H/V}}$ are almost degenerate~\cite{Gurioli2019,Saimon2021}. Nevertheless, a finite energy splitting (or fine-structure splitting, FSS) generally remains, leading to a time evolution of the entangled state and consequent entanglement degradation in time-averaged measurements~\cite{Stevenson2008}. Therefore, we make use of a piezoelectric strain-tuning device~\cite{Trotta2015,Huber2018} to cancel the FSS as shown in the right inset of \autoref{fig:fig1}(a). 
For the optical excitation of an individual QD, resonant two-photon excitation (TPE) [left inset of \autoref{fig:fig1}(a)] is used by tuning the energy $E_L$ of a pulsed laser with $\SI{80}{MHz}$ repetition rate to the half of the XX transition energy~\cite{Jayakumar2013,Mueller2014_tpe} and by setting the laser power to obtain maximum XX intensity ($\pi$-pulse conditions).
The recorded spectrum under TPE at $T=\SI{64.4}{K}$ is shown in \autoref{fig:fig1}(a). Besides the XX at $\SI{1.5761}{eV}$ and X at $\SI{1.5799}{eV}$, a weaker line at $\SI{1.5836}{eV}$ is also visible, which we denote as X$^*$ and attribute to a thermally populated excitonic state. Spectra collected at different temperatures, showing also further excited states, can be found in the Supplementary Material.\cite{sup}

To assess the effect of $T$ on the light emission characteristics and entangled photon generation following optical excitation we performed our study by increasing $T$ step-wise, in a range from \SI{4.4}{K} to \SI{64.4}{K}. For each $T$ value we cancelled the FSS via strain-tuning. First, the $g^{(2)}$ auto-correlation functions of both the XX and X were recorded. In \autoref{fig:fig1}(b) a clear broadening of the X histogram peaks is visible at  $T=\SI{64.4}{K}$ compared to low temperature data. In addition, a slight increase in $g^{(2)}(0)$ (see Fig. 3 Supplementary\cite{sup}) for XX and X from $g^{(2)}_{\mathrm{XX}}(0)=0.008(1)$ to $g^{(2)}_{\mathrm{XX}}(0)=0.032(5)$ and $g^{(2)}_{\mathrm{X}}(0)=0.008(1)$ to $g^{(2)}_{\mathrm{X}}(0)=0.033(3)$ is visible.

A full state tomography was performed to obtain the two-qubit density matrices in polarization space, using a maximum likelihood method~\cite{JamesKwiatPRA2001}. Two representative density matrices for $\SI{4.4}{K}$ and $\SI{48.4}{K}$ are shown in \autoref{fig:fig1}(c) and (d). For higher temperatures, we observe decreasing VV-HH coherence with respect to HH-HH and VV-VV occupations, as well as rising HV and VH elements, indicative of state mixing. We evaluate concurrence and fidelity at every temperature for a time bin of $\SI{2}{ns}$, as shown in \autoref{fig:fig1}(e). At low temperatures the concurrence [fidelity] is equal to $0.94(1)$ [$0.969(4)$], 
comparable to former investigations~\cite{Huber2018,Schimpf2021_20K, Hafenbrak2007}. The degree of entanglement stays approximately constant up to about $\SI{15}{K}$ and then decreases with increasing temperature. 
The slight increase in $g^{(2)}(0)$ mentioned above does not explain the observed steep entanglement degradation shown in \autoref{fig:fig1}(c-e) since the $g^{(2)}(0)$ values are still in the range typically observed at $T=\SI{5}{K}$\cite{Reindl2019,Schimpf2021_20K,keil2017}.

%
%
To gain further insights in the origin of the entanglement degradation, we study the decay dynamics of the XX and X emission following excitation as shown in \autoref{fig:fig2}. The upper half of \autoref{fig:fig2}(a) shows the normalized measured decay traces for all explored temperatures. 
%
At $T=\SI{4.4}{K}$, the XX emission shows an exponential decay, from which we extract a XX lifetime of $\SI{129(3)}{ps}$. 
With increasing $T$ the XX decay first shows an increased slope and then becomes slower at longer timescales, resulting in a crossing of the $\SI{4.4}{K}$ and the $\SI{64.4}{K}$ decay traces at around $\SI{0.6}{ns}$. 
A much stronger $T$-dependence is observed for the X decay: At $T=\SI{4.4}{K}$ the X intensity first rises and then exponentially decays with a time constant of $\SI{231(4)}{ps}$, as expected for a cascaded decay. For increasing $T$ we see a slightly accelerated decay at short time scales similar to XX and -- most importantly -- an increasingly pronounced tail for long timescales, reminiscent of the slow decay observed under excited state excitation at low temperatures~\cite{Reindl2019, Jahn2015}.
%
To understand this behavior, we extend the exciton-decay model of Ref.~\citenum{Tighineanu2013} to the biexciton decay. On top of the usual four-levels shown in the inset of \autoref{fig:fig1}(a), i.e. the ground state $\ket{G}$, the bright exciton states $\ket{X_{H/V}}$ and biexciton $\ket{XX}$, we add two dark exciton states $\ket{X_D}$, as well as excited exciton $\ket{X^*}$ and excited biexciton states $\ket{XX^*}$, see \autoref{fig:fig2}(b). In the single particle picture, the ``hot'' $\ket{XX^*}$ and $\ket{X^*}$ states are configurations where the electrons are in the ``$s$-shell'' and the holes in the ``$p$-shell''. The energy difference $\Delta E=\SI{3.7}{meV}$ between $\ket{X}$ and $\ket{X^*}$ is taken from the recorded spectrum in \autoref{fig:fig1}(a).
Because $\ket{X^*}$ consists of an electron and a hole, four spin configurations are possible, resulting in four possible transitions. For purely heavy-hole excitons we would expect two bright and two dark states, similar to the ground state exciton. In Ref.~\citenum{HuberLehner2019} a triplet was instead observed, which we ascribe to the high light-hole contribution of almost $40 \%$ (with $\sim 25 \%$ of bright admixture) to ``$p$-shell'' holes~\cite{HuberLehner2019,Huo:NatPhys,Belhadj2010}. As we are not interested in the detailed population of the excited states, we include $\ket{X^*}$ as a single state with multiplicity of four. 
For the $\ket{XX^*}$, we expect two ``$s$-shell'' electrons in a singlet state and two holes, one in the ``$s$-shell'' and the other in the ``$p$-shell'', resulting again in four possible states. Since we were not able to unequivocally identify the emission lines associated with $\ket{XX^*}$, 
we assume the same value of $\SI{3.7}{meV}$ for the $\ket{XX}-\ket{XX^*}$ energy separation. 
Finally for the bright-dark splitting we take a value of $\SI{110}{\micro eV}$, as in Ref.~\citenum{HuberLehner2019} and a multiplicity of two.
We note that excited states with the electron in the ``$p$-shell'' and the hole in the ``$s$-shell'' are unlikely to be populated in the explored temperature range due to significantly higher energy differences of 15--20~\si{\milli\electronvolt} according to our calculation (see Supplementary Materials\cite{sup}). States with holes in higher-energy shells play instead a role for $T\gtrsim\SI{40}{K}$ (see below), but are omitted from our model for the sake of simplicity. 

%

With reference to the levels shown \autoref{fig:fig2}(b) we now focus on the radiative recombinations (solid lines) and non-radiative phonon-assisted transitions and their rates. Different from the dominant XX and X emission lines, which are characterized by the rates $\gamma_X, \gamma_{XX}$, the recombination rate $\gamma_{X^*}$ of $\ket{X^*}$ is relatively weak because of the different envelope function symmetry for the electron and hole but clearly visible at high temperature [\autoref{fig:fig1}(a)], under non-resonant excitation\cite{HuberLehner2019}, and in photoluminescence (PL) excitation measurements~\cite{Reindl2019,Rastelli2004a}. For radiative recombinations involving the same single-particle states we assume the same values for the corresponding rates. As an example, the recombination of $\ket{XX^*}$ leaving the system in a ground state exciton $\ket{X_{H/V}}$ ($\ket{X_D}$) takes place with a rate $\gamma_{XX_1^*}$ ($\gamma_{XX_D^*}$) with $\gamma_{XX_1^*}=\gamma_{XX_D^*}=\gamma_{X^*}$, since in all cases we have a ``$p$-shell'' hole recombining with an ``$s$-shell'' electron. Further, the rate $\gamma_{XX_2^*}$ for the $\ket{XX^*}$ recombination leaving the system in the $\ket{X^*}$ state is assumed to be equal to $\gamma_{X}$ since the ``$s$-shell'' electron and ``$s$-shell'' hole are more likely to recombine first.

For the phonon-assisted transitions we assume a temperature dependence that is determined by the expected number of phonons excited at temperature $T$ according to the Bose-Einstein distribution: 
\begin{align}
    n(\Delta E, T)=\left(\exp\left[\frac{\Delta E}{k_B T}\right]-1\right)^{-1},
\end{align}
where $k_B$ is the Boltzmann constant. For $T\simeq\SI{43}{K}$, $k_B T$ is comparable to $\Delta E$, making the occupation of hole-dominated hot states likely. The excited exciton state $\ket{X^*}$  relaxes to $\ket{X_{V/H}}$ with a rate:
\begin{align}
    \gamma_{PH}&=\gamma_{PH}^0\left[1+n(\Delta E,T)\right],\label{eq:eq_gammaPH}
\end{align}
where $\gamma_{PH}^0=\SI{1}{ns^{-1}}$ 
is the phonon-assisted relaxation rate at low temperature, estimated from simulations (see Sec. Phonon-assisted relaxation of the Supplementary Material\cite{sup}) and fully consistent with the slow relaxation previously reported in Refs.~\citenum{Reindl2019} and \citenum{Jahn2015} for similar QDs.
$\ket{X^*}$ is populated via the phonon mediated rate $4\gamma_{PH}^*$, where the factor 4 in this and other rates corresponds to the state multiplicity discussed above, with:
\begin{align}
    \gamma_{PH}^*&=\gamma_{PH}^0 n(\Delta E,T)=\text{exp}\left[-\frac{\Delta E}{k_B T}\right]\gamma_{PH}
    \label{eq:eq_gammaPHs}
\end{align}

%
To justify the assumption of equal phonon-assisted rates for transitions involving states with different spin configurations and to fully understand the entanglement degradation observed in \autoref{fig:fig1}(e), it is important to note that the highly mixed character of excited hole levels implies that the spin projection along the growth axis is not a good quantum number for the hot states.
As a consequence, phonon-assisted relaxations are effectively not spin-conserving (see Supplementary Material\cite{sup}). This is in good agreement with PL-excitation measurements, where the $\ket{X^*}$ was excited resonantly and equal intensities for the $X_V$ and $X_H$ transitions independent of laser polarization were observed after relaxation. Calculations also confirm this almost ``spin-agnostic'' relaxation from all four $\ket{X^*}$ to all bright and dark $\ket{X}$ states with at most 40~\% difference in rates (see Supplementary Material\cite{sup}), which justifies taking them equal for the sake of simplicity. The same approach is followed for the transitions involving  $\ket{XX^*}$.
Note that spin-flips between bright and dark exciton states are neglected since they are expected to occur at timescales in the order of $\si{\micro s}$~\cite{Tighineanu2013}. All rates used in our model and the corresponding origin are summarized in \autoref{tab:tab1}. 

\begin{table}
  \begin{threeparttable}
    \caption{Rates of the decay model used for theoretical calculations.}
    \renewcommand{\arraystretch}{1.2}
    \label{tab:tab1}
     \begin{tabular}{p{0.1\columnwidth}|p{0.8\columnwidth}}
    \textbf{Rate}  & \textbf{in ns\textsuperscript{-1}}  \\  \hline\hline
        $\gamma_X$ & 1/0.231(4)\textsuperscript{a} \\ \hline
        $\gamma_{X^*}$ & 1/10\textsuperscript{b}\\ \hline
        $\gamma_{XX}$ & 1/0.129(3)\textsuperscript{a} \\ \hline
        $\gamma_{XX_D^*}$ & $=\gamma_{X^*}$ \\ \hline
        $\gamma_{XX_1^*}$ & $=\gamma_{X^*}$ \\ \hline
        $\gamma_{XX_2^*}$ & $=\gamma_X$  \\ \hline
        $\gamma_{PH}^0$ & 1\textsuperscript{c}  \\ \hline
     \end{tabular}
    \begin{tablenotes}
      \small
      \item \textsuperscript{a}Values taken from the PL measurements following TPE excitation at $\SI{4.4}{K}$.
      \item \textsuperscript{b}Value estimated from the comparison between the results of the rate equation model and PL intensities of the X$^*$ line as well as from the $k\cdot p$ and CI simulations (see Supplementary Material\cite{sup}).
      \item \textsuperscript{c}Values estimated from the $k\cdot p$ and CI simulations.
    \end{tablenotes}
  \end{threeparttable}
\end{table}

\begin{figure*}[htb]
	\includegraphics[width=1\textwidth]{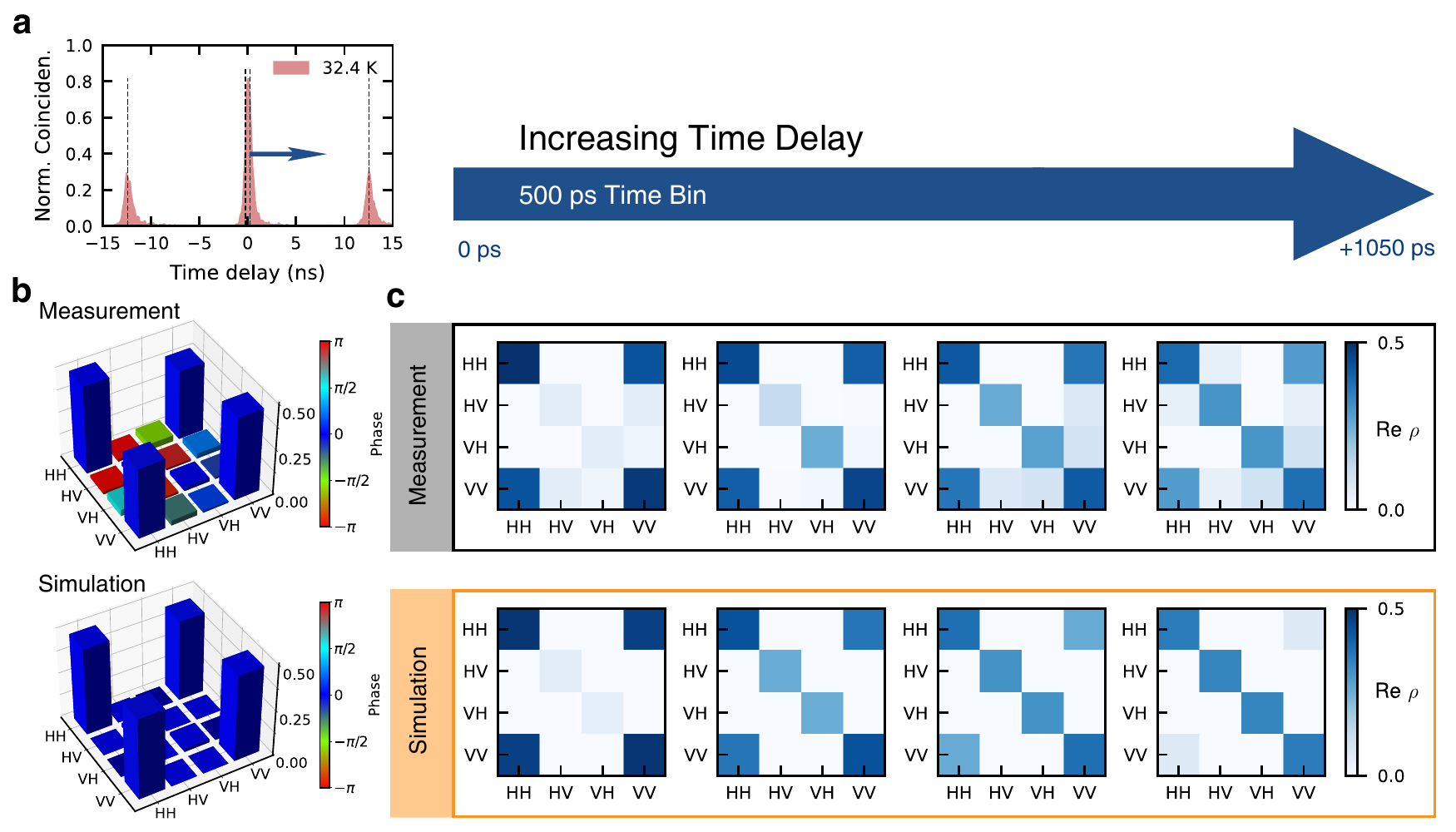}
	\caption{Comparison of measured and calculated two-photon density matrices at a temperature of 32.4 K for a time bin of $\SI{500}{ps}$ and different bin center. (a) Exemplary coincidence histogram with a broadening on the right side of each peak stemming from slow X decay. Black dashed lines show the time bin used to calculate the density matrices shown in (b). The blue arrow shows the direction in which the time bin window is moved for time filtering beginning at zero time delay with the bin shown in (a). (c) 2D representations of the real parts of the density matrices with increasing time delays in equidistant steps, as obtained from the experiment (grey box) and theory (orange box). The leftmost matrices correspond to the time bin shown in (a) and used for (b).}
	\label{fig:fig3}
\end{figure*}
Solving the rate equations for the presented system and convolving the obtained time evolutions with a measured instrument response function results in the decay traces shown in the lower half of \autoref{fig:fig2}(a). The simulation reproduces both the initial acceleration of the decay observed for the XX and X -- due to population loss through thermal excitation of the hot states -- and the pronounced slow decay of the X signal -- due to the re-population of the bright $\ket{X_{V/H}}$ states. As in the experiment, the operation temperature mostly affects the X decay. 
However, we notice that, at higher temperatures, the measured X decay is still slower than the predicted decay. We attribute this to additional phonon-assisted excitation to higher-energy hot states at temperatures $>\SI{40}{K}$, which were not taken into account in the presented model (see Supplementary Material\cite{sup}).

We now turn to the effect of the thermally-induced processes described above on the degree of entanglement of photon pairs generated by the biexciton-exciton decay cascade. To this end, numerical simulations are performed, solving the corresponding Liouville-von Neumann equation. In a first step, the photonic two-qubit density matrix is theoretically calculated based on polarization-resolved, time-integrated two-time correlation functions\cite{Seidelmann2022}, modelling the measurements performed in the experiment. Afterwards the simulated concurrence is directly evaluated from the obtained photonic density matrices.
The result of these simulations is shown in \autoref{fig:fig1}(e) in light blue for temperatures from $0$ to $\SI{70}{K}$.
Note that the simulations do not predict a unity concurrence for low temperatures even for zero fine-structure splitting, since the TPE sets a limit to the obtainable degree of entanglement due to a dynamic Stark shift of one exciton level induced by the excitation itself\cite{Seidelmann2022}. The simulated result reproduces well the concurrence plateau at lower temperatures followed by a decrease starting around $\SI{16}{K}$. Again, theory and experiment are in good agreement up to a temperature of $\SI{40}{K}$. For higher temperatures, the theory predicts a slower degradation compared to the experiment, consistent to additional phonon-assisted excitation channels. 

In addition to the concurrence calculations, we use the decay model shown in \autoref{fig:fig2}(b) to compute the photonic two-qubit density matrices and compare them with the experimentally reconstructed matrices, where mixing and decoherence emerge with increasing temperature, see \autoref{fig:fig1}(d).
In \autoref{fig:fig3}(a) an exemplary XX-X coincidence histogram measured at a temperature of $\SI{32.4}{K}$ is shown, displaying coincidences arising from the slow X decay on the right side of the peaks. Coincidences within the chosen time bin of $\SI{500}{ps}$ (black dashed lines), corresponding approximately to our detector resolution are summed up and compared with the average area of the side peaks in the same interval. 
\autoref{fig:fig3}(b) shows the measured and simulated real part of the density matrices for this time bin. The leftmost 2D diagrams in the grey and orange boxes in \autoref{fig:fig3}(c) correspond to the 3D representations in \autoref{fig:fig3}(b). Next, we begin to shift the time bin to higher time delays, indicated by the blue arrows in \autoref{fig:fig3}(a) and above \autoref{fig:fig3}(c). The resulting density matrices for a time delay up to $\SI{1050}{ps}$ are shown in panel (c). The grey box shows the measurement, the orange box the density matrices obtained from corresponding simulations. In order to mimic the time-filtering analysis theoretically, only photon pairs with a delay time in the respective interval/time bin are considered in the time-integrated correlation functions, cf., Supplementary Material\cite{sup}.
Both, experiment and theory show increasing state mixing (rise of HV-HV and VH-VH elements)  and decoherence (drop of HH-VV and VV-HH elements) with increasing time delay. In turn, this finding is consistent with our dynamic model, in which the detection events producing the ``tail'' in the coincidence histograms are ascribed to thermal cycling among levels, i.e. the occupation of hot- and dark-states at elevated temperatures followed by bright exciton re-population with no spin memory. 

\section{Discussion and Outlook}
We have investigated the effect of rising operation temperature on the quality of the polarization entanglement of photon pairs generated by the biexciton-exciton decay cascade in a single GaAs QD tuned to have negligible excitonic fine-structure splitting. By performing full state tomography including time-filtering and lifetime measurements under resonant optical excitation as well as dedicated calculations, we ascribe both the entanglement degradation and changes in decay dynamics to thermal cycling among the desired $\ket{XX}$ and $\ket{X}$ states and ``undesired'' hot and dark states, which are connected to the former by phonon-assisted transitions leading to spin scattering and decoherence. In turn, the spin-agnostic character of the transitions is traced back to the high valence band mixing in the excited states of the employed QDs.

From the achieved understanding one could envision that an increased energy splitting $\Delta E$ can substantially extend the plateau of high concurrence at lower temperatures up to the operation temperatures of $\SI{40}{K}$, reachable with available Stirling coolers \cite{CryoBook,Schlehahn2015}. Since the excited states will be less populated for larger energy splittings, the impact of thermal cycling is reduced. An increasing $\Delta E$ is also expected to lead to a reduction of hole mixing in the excited states benefiting the preservation of high entanglement. Consequently, we anticipate that QDs capable of generating highly entangled photons with relaxed operation-temperature requirements can be obtained by slightly reducing the QD size. For GaAs QDs, this can be simply achieved by reducing the amount of GaAs filling~\cite{Wang2009a}. However, along with a change in QD size one must consider also pure dephasing~\cite{IlesSmith2017} due to the deformation potential coupling to longitudinal acoustic (LA) phonons~\cite{Besombes2001,Krummheuer2002,RamsayPRL104}. This mechanism has been shown to reduce the concurrence by enhancing off-resonant single-photon transitions and decoherence~\cite{Seidelmann2019}. Although the effects of pure dephasing seem to be insignificant for the QDs studies in this work, they become more relevant with decreasing QD size as the coupling to LA phonons is more effective~\cite{GlaesslPRB2011,LuekerPRB2017}.

Nevertheless, recent work~\cite{Mueller2018} on (presumably strongly-confining) InGaAs QDs shows indeed the persistence of high levels of entanglement beyond 90~K. In conclusion we can say that finding the optimal size for LDE-grown GaAs QDs to tailor the entanglement at elevated temperatures needs a thorough understanding of various effects. Within the presented work we could identify the evolution of the decay dynamics and the loss of spin-memory due to the occupation of excited states as major consequences of temperature.
\section{Data availability}
The data of this study is available from the corresponding author upon request.

\begin{acknowledgement}
We thank G.~Weihs, M.~Reindl, J.~Freund, M.~Peter and T.~M.~Krieger for fruitful discussions.
M.~G. is grateful to K.~Gawarecki for sharing his $\bm{k}\cdot\bm{p}$ code.
This work was financially supported by the Austrian Science Fund (FWF) via the Research Group FG5, P 29603, P 30459, I 4320, I 4380, I 3762, the European Union’s Horizon 2020 research and innovation program under Grant Agreements No. 891366 (QD-E-QKD),  No. 899814 (Qurope) and No. 871130 (Ascent+), the Linz Institute of Technology (LIT), and the LIT Secure and Correct Systems Lab, supported by the State of Upper Austria.
T.~S. and V.~M.~A. are grateful for support by the Deutsche Forschungsgemeinschaft (DFG, German Research Foundation) via the project No.~419036043. X.~Y. acknowledges support from National Natural Science Foundation of China (NSFC 12104090). D.~E.~R. acknowledges support by the DFG via project number 428026575 (AEQuDot). 
Part of the calculations have been carried out using resources provided by Wroclaw Centre for Networking and Supercomputing (http://wcss.pl), Grant No. 203.


\end{acknowledgement}

\begin{suppinfo}
Supporting Information (SM)\cite{sup}: Additional experimental details, materials and methods. SM includes new references. \cite{BahderPRB1990,GawareckiPRB2014,Mielnik-PyszczorskiPRB2018,GawelczykPRB2017,Andrzejewski_2010,ThranhardtPRB2002,KrzykowskiPRB2020,WoodsPRB2004,Lindblad:1976,Wootters1998}
\end{suppinfo}
\bibliography{references}

\end{document}